# Perspectives on a 6G Architecture

Rainer Liebhart, Mansoor Shafi (Life Fellow, IEEE), Gajan Shivanandan (Member, IEEE), Devaki Chandramouli (Sr. Member, IEEE), Laurent Thiebaut

*Abstract*
Mobile communications have been undergoing a generational change every ten years. Whilst we are just beginning to roll out 5G networks, significant efforts are planned to standardize 6G that is expected to be commercially introduced by 2030. This paper looks at the use cases for 6G and their impact on the network architecture to meet the anticipated performance requirements. The new architecture is based on integrating various network functions in virtual cloud environments, leveraging the advancement of artificial intelligence in all domains, integrating different sub-networks constituting the 6G system, and on enhanced means of exposing data and services to third parties.

## I. INTRODUCTION

Wireless communication is an integral part of our daily lives and the social fabric. Applications connecting humans to machines, and machines to machines are fuelling explosive growth. The number of connected machines could become 59 times larger than the world population [9]. In response, the mobile industry has been experiencing a generation change almost every 10 years. Whilst 5G is still being deployed and 5G Advanced is being developed, efforts are underway to research and eventually standardize 6G (IMT 2030) wireless systems that are likely to be introduced from 2030 onwards. Recently many research papers and white papers have appeared describing the vision of 6G, performance requirements, enabling technologies, and architectural principles [1], [2], [7]-[9], [12]. IEEE conferences ICC and Globecom have devoted many sessions and panel discussions to 6G. ITU-R WP 5D is in the process of preparing a new recommendation for the vision of IMT 2030 [3] to be followed by a report on minimum requirements for IMT 2030. A recommendation on future technology trends [10] has recently been approved. Work is also underway to identify new spectrum bands for IMT 2030 to be agreed at WRC 27. IMT 2030 will operate in all existing bands ranging from sub 1 GHz, mid bands (3-6 GHz) and mmWave bands 24-100 GHz. As a consequence of the exponential data growth also new bands are needed: 7-24 GHz and 100-300 GHz. The 3GPP is currently focusing on Release 18 (5G Advanced) and overtime will move to standardize 6G.

The 6G system will support a wide range of use cases (to be discussed later) and will support diverse and integrated network topologies including satellite and terrestrial systems, drones, aircraft, personal networks, etc. Whilst it is premature to describe the 6G architecture right now, we will investigate the principles to be adopted in the development of a new 6G network architecture in this paper. The format of this paper is as follows: A brief overview of 6G services and their key performance indicators is given in section II. A recap of the 5G core network architecture is provided in section III. Motivation, drivers, and perspectives for a 6G architecture are outlined in section IV, followed by conclusions in section V.

## II. OVERVIEW OF 6G SERVICES AND KPIS

This section gives an overview of 6G services and their proposed KPIs. In [2] and [3], the use cases for IMT 2030 are categorized as:

**Immersive Media:** Extended reality (XR) encompasses VR (Virtual reality), AR (Artificial reality) and MR (Mixed Reality) and is of increasing interest in entertainment, medicine, education, and manufacturing. Truly immersive AR, XR media streaming, 16K UHD video require about 0.44 Gbps throughput [9].

**High Fidelity Holographic Society:** Advances in high resolution rendering and wearable technologies could make holographic telepresence the mode of choice for future communication, including multi-model communication for teleoperation, haptic and the tactile Internet. Data rates for holograms are more than 1 Tbps ([2], [9]).

**Enhanced Machine Type Communication:** One important use case for 6G is machine to machine communication. The performance targets for this kind of communication well exceed the current targets for human to human or human to machine communication, requiring e.g., a huge connection density.

**Enabling new services:** These include extreme accurate positioning, localization and tracking in 3D, interactive mapping, digital twins and virtual worlds, digital health care, smart industry, automatic detection protection and investigation.

**Enhanced broadband:** Wireless access points in metro stations, shopping malls, and other public places will be deployed as part of ultra-dense radio networks and providing extreme high data rates, low latency, and high positioning accuracy.

Table I provides some selected KPI metrics (relevant to the subject of this paper) and their values, being discussed in [3]. In addition, we expect new KPIs for "trust", "security" and "sustainability". This is because the 6G use cases pose new security challenges, due to the increased points of vulnerability created by the large number of devices connected to 6G networks and the upcoming capabilities of quantum computing. Additionally, the use of 6G in many applications involves mission critical infrastructure – in turn this requires the 6G infrastructure to be dependable, resilient, and attack resistant. Trust encompasses privacy and resilience. Privacy preservation is essential to realise the full potential of IMT 2030 use cases. Privacy preserving techniques for IMT 2030 are discussed in [6]. The evidence of a service and the underlying network to meet expectations about reliability, security, safety, and availability will help to build trustworthiness. As far as security and privacy are concerned, the IMT 2030 network should be responsive to user initiated and programmable actions to ensure protection of the personal information stored in the cloud and exposed to various computational systems.

TABLE I
KPI VALUES OF SELECTED METRICS FOR 6G

| KPI | Definition | Proposed Value |
|---|---|---|
| Peak Down link data rate | Maximum achievable data rate under ideal conditions per user when all the radio resources are allocated to the down link. | Up to 1 Tbps needed for holographic communication, VR/AR, tactile internet applications and extremely high rate information showers. |
| Per user experienced rate | Achievable data rate that is available ubiquitously across the coverage area to a mobile user/device- typically at the 5% cdf level. | 1 Gbps |
| Connection density | Total number of connected and/or accessible devices per unit area (per km$^2$). | $10^7$ devices/km$^2$ Given the desire for IMT-2030 systems to support an internet-of-everything, the connection density is expected to be 10× that of 5G. |
| Reliability | The capability of transmitting a given amount of traffic within a predetermined time duration with high success probability. | $1\text{-}10^{-7}$ to $1\text{-}10^{-9}$ |
| User Plane Latency | The time from when the radio source sends a packet to when the destination receives it. | 10 μs to 1 ms Holographic, VR/AR and tactile applications |
| Control Plane Latency | | 20 ms |

A quick glance at the KPI values in Table I reveals that the corresponding values of the metrics are truly formidable and challenging. A flavour of the challenges is discussed in [2]. Take for instance the peak rate of up to Tbps, this will require very large carrier bandwidths. Such large carrier bandwidths are not available in the currently allocated 5G bands. There is significant interest to explore the sub-THz bands in the range of 140 to 350 GHz for 6G. However, IMT 2030 spectrum does not depend on the availability of sub-THz bands. Not all the IMT 2030 applications will be suitable for these bands which are constrained by a limited range in addition to the challenges in providing sufficient transport connectivity. Therefore, 6G spectrum must become available at all frequency ranges, including sub 1 GHz bands, mid-bands like 7-24 GHz, and mmWave bands above 24 GHz. Providing sub ms latency requires an optimized physical layer, but also improvements at higher layers. These extremely low latency requirements cannot be easily realized by the current transport and mobile core networks. Therefore, a significant reduction of complexity in the current transport and mobile core networks seems necessary. The 6G network needs to cope with enhanced requirements regarding delivery of packets, e.g., delivering packets within a very short time window. In case of an application that consists of traffic flows from different sources (say distributed orchestra), all components must arrive within a specified time window to avoid the feeling of cyber sickness. All of this does not only lead to an absolute reduction of latency but also to a new kind of in-built flexibility considering latency and jitter.

## III. THE 5G SYSTEM ARCHITECTURE

3GPP Release 15 specified a functional architecture for the 5G system [5], Figure 1, which comprises of the User Equipment (UE), the 5G Radio Access Network (5G RAN) and the 5G Core (5GC). The 5G System consist of so-called Network Functions (NF) which are the smallest set of functionalities deployable in a multi-vendor environment. Some key principles of the new 5G architecture are the following: support multiple access technologies, separation of control and user plane, separation of compute and storage capabilities of a NF, service-based architecture (SBA) and modularization of the functional and interface design.

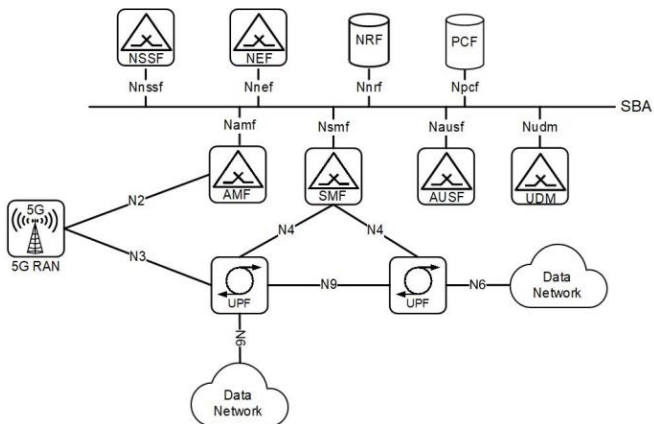

**Figure 1:** 5G architecture

The SBA as defined in [5], represented a major evolution towards an architecture based on web scale type of technologies. In SBA networks, functions are implemented as software defined services where each service can be used by one or more consumers; more details are in [11]. This allows functions that make up the 5GC to be hosted in a cloud native environment where NFs are deployed either as Virtualized Network Functions (VNF) or Containerized Network Functions (CNF). While dedicated hardware deployment options are not precluded by 3GPP specifications, in practice all implementations have followed the cloud native based approach which enables a high degree of modularity, reusability, and self-containment of NFs. The disaggregation of the functions within the 5GC lends itself to operators taking advantage of a wide range of deployment architectures spanning from centralized to distributed deployments depending on the use case and further disaggregation in 6G architectures. The major 5G network functions are the Access and Mobility Management (AMF), the Session Management (SMF), the User Plane Function (UPF) and the User Data Management (UDM) function. Besides these basic functions, fundamental to the 5G architecture was the flexibility introduced by advanced network functions, such as the Network Exposure Function (NEF) and Network Slicing Selection Function (NSSF) allowing operators to introduce more innovative services and exposure data to third parties. Another key use case enabled by the 5GC, is the support of Mobile Edge Computing (MEC) based deployments. The 5G system allows core network functions (such as UPF) and applications to be placed at the edge of the network while enabling devices to be mobile when using such applications. The 5G system is evolving in 3GPP Releases 18 and 19 towards 5G Advanced. With 5G Advanced new features and capabilities are introduced to cater with future demands of mobile networks and will pave the path towards 6G.

## IV. MOTIVATION, DRIVERS, AND CONCEPTS FOR A 6G ARCHITECTURE

The use cases of 6G and their KPIs cannot be efficiently provided by the 5G System (see [7]-[9]). In 5GS, reduction of latency is achieved by radio layer improvements and moving computing capability closer to the mobile device. The ITU-T FG 2030 has published a series of reports on the network 2030, indicating the gaps in present networks to realize 6G use cases and considerations for a new network architecture [4]. In 6G the use cases will require a true convergence of communication and computing, enabling the user's device to use computing power in the network efficiently. The 6G network will support a wide variety of sub-networks consisting of fixed base stations, base stations that are moving, non-terrestrial base stations in various altitudes, different air mobility systems etc. These types of sub-networks are an integral part of the 6G architecture, allowing for seamless mobility between different coverage layers. We anticipate following main principles a future 6G network will follow:
- Adopting service-based architecture principles to the Radio Access Network (RAN)
- Enabling application aware Quality of Service (QoS)
- Integration of improved and new sustainability metrics
- Extended use of Artificial Intelligence (AI) and Machine Learning (ML)
- Introducing the concept of "Network of Networks" including a densified RAN
- Adopting new security principles
- Migration strategies

In the following we provide a closer look into these key technology areas. The main drivers, and potential concepts for a 6G architecture are followed by a high-level overview of the 6G architecture.

PROSPECTS OF A 6G ARCHITECTURE

At this stage it is premature to provide a functional architecture of a future 6GC like in Figure 1.

Basic decisions have not been made so far: what is the functional split inside the 6G Radio Access Network (RAN), inside the 6G Core Network (CN) and between radio and core networks. Adopting concepts such as SBA to the RAN and to the RAN-CN interface might be natural (see the following section) but the functional and protocol details and their potential consequences need to be carefully studied before making any decisions and moving into the standardization phase. However, some principles a 6G system must be based on seem to be clear:
- Supporting fully virtualised and non-virtualised network functions and components.
- Supporting applications requiring integration of heterogenous cloud environments: Far Edge, Edge, Regional and Central Clouds (Figure 2).
- Supporting best effort and QoS aware applications and high precision services requiring bandwidths ranging from Gbps to Tbps and extreme low latency for real time applications.
- Integration of heterogenous (sub-)networks.
- Advancements in the integration of computing, routing, and storage capabilities.
- Enhanced means of exposing data and services to third parties.
- Leveraging "as-a-Service (aaS)" like concepts such as Network as a Service, Platform as a Service, and Infrastructure as a Service.

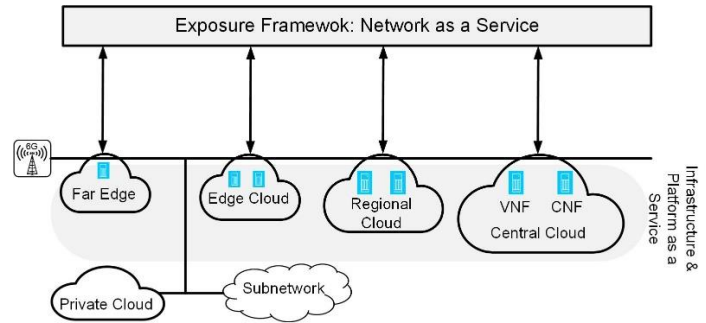

**Figure 2:** Heterogenous cloud deployments in 6G

SERVICE-BASED ARCHITECTURE IN RADIO ACCESS NETWORK AND CORE NETWORK

Mobile networks are defined by functionalities divided into Core Network (CN) and Radio Access Network (RAN) which are connected through open, multi-vendor interfaces. A key architectural change in 5GC was the transition to a cloud native service-based approach. This trend will further extend towards the edge and the radio access in 6G with the benefit of enabling end-to-end deployments using a harmonized, cloud-based framework with common operational tools. In comparison to 5G, the forthcoming 6G network needs to accommodate to an even more diverse set of use cases, services, and access technologies with different topologies. Cloud-based service delivery platforms will be diversified into on premises, edge, core and public clouds through different processing and service capabilities matching the needs of these services. The target architecture for 6G should be flexible enough so that network functions and services can be deployed across different cloud platforms, edge, and central cloud sites. Edge sites will host time sensitive radio related, e.g., Distributed Unit (DU), processing functions that benefit from HW acceleration. However, the flexibility of locating lower layer radio processing is limited by delay and capacity constraints, creating high demands on the transport network. The regional cloud hosts less time sensitive functions e.g., the Central Unit (CU) implementing higher layers of the radio protocol stack and core functions like distributed UPF or SMF. As a consequence of the transition to a cloud-native service-based approach in the CN and in future in RAN, different functional splits between CN and RAN can be investigated in the 6G scope. This could finally lead to an architecture where different functions that are traditionally located in CN and RAN are collocated in one function (Figure 3). Such collocated functionality can be part of the control plane (CP) or user plane (UP). This will allow for higher deployment flexibility, e.g., flexible placement of signalling and user plane functions close or far away from the radio site depending on economic needs and application demands.

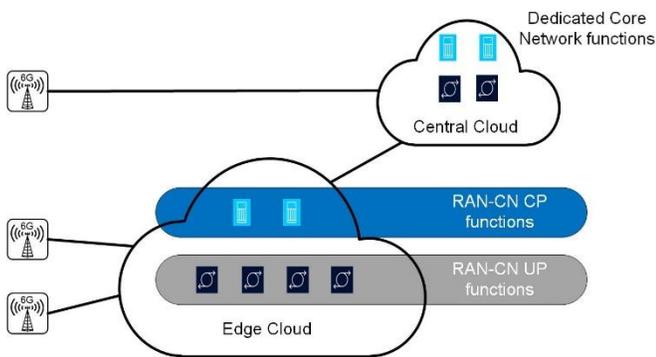

**Figure 3:** RAN-Core CP and UP functions

## APPLICATION AWARE QOS

In future a tighter coordination between applications and communication systems must be leveraged to improve end user experience. Real time immersive services are quite sensitive to packet loss, data rate and delay fluctuations. The need to support extreme throughput, real-time services such as holographic and Mixed-Reality (MR) communications, coupled with stricter security requirements, requires mobile networks to be aware of the QoS needs of applications. Low latency applications may also suffer from latency increase due to congestion that can occur either because of sudden link capacity changes or greedy traffic flows sharing the same link. Although real-time applications can adapt the transmission rate to sustain Quality of Experience (QoE), they are driving "blind". The current concept of guaranteed bit rates offered in cellular networks can run into scalability issues as it consumes lot of resources to sustain high data rates and low latency. To improve scalability, adaptive bit rate mechanisms and application aware QoS can be considered to effectively use resources.

## SUSTAINABILITY

Sustainability is a major requirement to cope with environmental challenges. The 6G architecture must map to UN's sustainable development goals [7] to improve life quality, enabling healthcare, learning, reduce carbon footprint, maximize handprint and eliminate digital divide. Thus, sustainability and energy efficiency are goals that need to be looked at from a holistic point of view in the 6G era. For 6G, the overall objective is to design an end-to-end architecture and protocols enabling an optimized, scalable and energy efficient operation of the whole network, under the assumption that most or all network functions are virtualized and deployed in Edge or Central data centres.

## AI AND ML

Artificial Intelligence (AI) and Machine Learning (ML) technologies will be essential ingredients of 6G. They will have a huge impact on all domains in the mobile network. Support of local AI, joint AI between two domains or end-to-end AI is required. These technologies provide new means for network automation capabilities, boosted performance, and enhanced energy efficiency. Use of AI/ML inside 6G RAN and CN requires support of new functions and interfaces enabling data collection, data processing, and data distribution at large scale while being cognizant of sustainability needs. In addition, training for model refinement and updating inference models within network functions is foreseen. AI-driven closed-loop automation enables automatic detection of complex network issues or fault situations, leading to an automated way for resolution of such issues, and implementing required changes in a highly automated framework. Another possible use case for AI/ML techniques in 6G is the optimization of RAN functions by means of adapting AI/ML concepts i.e., self-learning scheduling and beamforming algorithms considering data from radio sites and their neighbours to adapt and optimize the algorithms' behaviour.

## NETWORK OF NETWORKS

The 6G network will comprise of many heterogeneous, multi-domain sub-networks serving different purposes and use cases. These networks are connected to a single 6G Core. This concept is called a "Network of Networks" (NoN), see Figure 4 and [8].

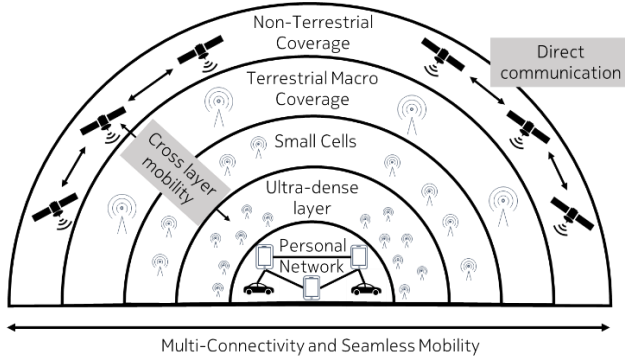

**Figure 4:** Network of Networks

Sub-networks in this context can be for example, wireline or wireless networks using different types of technologies such as copper, fibre, 3GPP cellular, Bluetooth and IEEE Wi-Fi and/or different frequency bands (low, mid, high, and ultra-high bands). Despite leveraging different access technologies, a sub-network can be deployed as public or private network, dedicated to specific use cases, e.g., for the Internet of Things, as a satellite or personal network. The goal of the NoN architecture is to hide this kind of heterogeneity and enable seamless service experience for human users and machines. The ultimate goal is to combine diverse access and deployment options to aggregate capacity, increase reliability, extend security, and allow for a seamless connectivity and interworking experience wherever and whenever a device connects via one or even multiple access networks simultaneously. Therefore, key components of a future 6G system will be multi-connectivity-based mobility and cognitive data planes to enable seamless inter-working between sub-networks.

## 6G SECURITY PRINCIPLES

Security will be an intrinsic part of the future 6G architecture. Some of the challenges a 6G security design needs to cope with are the high degree of AI/ML based automation and intelligence in the network, the deployment of multiple de-centralized (sub-)networks with customized security demands, and the ongoing trend towards virtualized networks. For example, data poisoning (i.e., an adversarial attack in planting malicious points during the training of ML algorithms due to weaknesses in the algorithms) creates a potential risk for ML based algorithms. In turn this may lead to false outcomes. Another risk is the potential leakage of user privacy related data due to the dissemination of these data required by AI/ML mechanisms. In addition, the evolution of quantum computing creates new threats and existing security algorithms may be replaced in future by quantum-safe mechanisms.

## MIGRATION FROM 5G TO 6G

As for previous generations, smooth migration from 5G to 6G will be essential. Spectrum is the first building block of wireless networks. Existing IMT bands should be utilized efficiently for 6G, however, 6G will also require extremely high bandwidth and wide band carriers (Figure 5). At the same time, 5G will be widely deployed on bands of different frequency ranges. Therefore, to cope with different coverage scenarios, various frequency bands need to be combined in 6G.

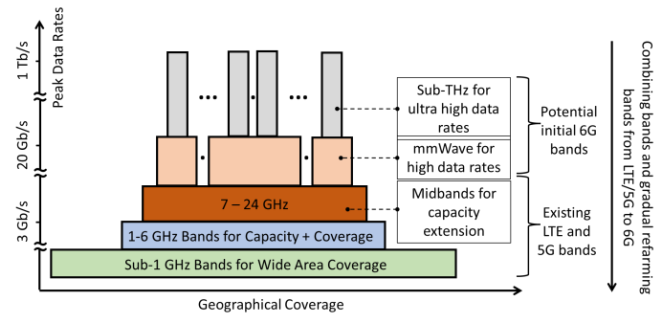

**Figure 5:** 6G deployment in multiple bands (data rates are taken from [2])

Efficient and gradual re-farming of spectrum from 5G to 6G will be required. Regardless of the concrete migration trajectory the end state is a 6G System supporting concepts as explained in previous sections of this paper.

## V. CONCLUSION

This paper described new use cases and applications anticipated for the upcoming 6G era. Some of these applications are extended reality, high fidelity holographic society, enhanced machine type communication, and extreme broadband. These services will require much higher reliability and throughput, and at the same time lower latency as supported today with 5G. The paper explained main pillars of a future 6G architecture which are in our view: adopting service based architecture principles to the Radio Access Network, enabling application aware QoS, integration of new sustainability and security measures, extended use of AI/ML and introducing new concepts like Network of Networks.

**Rainer Liebhart** has 30 years of experience in the telecommunication industry. He holds an MS in Mathematics from the Ludwig-Maximilians University in Munich, Germany. He works in Nokia as a research project manager and delegate in 3GPP SA and SA2. Rainer is (co-)author of around 100 patents in the telecommunication area, of several IEEE papers and co-editor of the books "LTE for Public Safety" (Wiley, 2015) and "5G for the Connected World" (Wiley, 2019).

**Mansoor Shafi** is a fellow at Spark New Zealand, Wellington, 6011, New Zealand. He has published widely in cellular communications and contributed to wireless standards in the International Telecommunication Union-Radiocommunication Sector as a New Zealand delegate for more than 30 years. He is an adjunct professor at Victoria University of Wellington and the University of Canterbury, New Zealand. He is a Life Fellow of IEEE and winner of multiple IEEE awards.

**Gajan Shivanandan** has over 20 years of experience in the cellular mobile industry across architecture and engineering roles. His current role is as End-to-end Architect (Mobile) for Spark NZ in Wellington. He has recently engaged with 3GPP standardization groups specifically 3GPP SA2 and RAN4. He holds a B.Tech in Information Engineering from Massey University in Palmerston North, New Zealand.

**Devaki Chandramouli** is a Nokia Bell Labs Fellow and Head of North American Standardization at Nokia. She serves as a Steering Group co-chair in the Next G Alliance. She is a delegate in 3GPP SA2. She has co-authored IEEE papers on 5G, co-edited a book on "LTE for Public Safety" (Wiley, 2015) and a book on "5G for the Connected World" (Wiley, 2019). She has co-authored over 230 patents in the area of wireless communications. Devaki received her B.E in Computer Science from Madras University (India) and M.S in Computer Science from University of Texas at Arlington (USA).

**Laurent Thiebaut** is the Head of the Nokia delegation at 3GPP SA2. He graduated from Ecole Nationale Supérieure des Telecommunications of Paris and has been working for 37 years in the Telco industry. Laurent is (co-) author of around 100 patents in the telecommunication area and contributed to the book "5G for the Connected World" (Wiley, 2019).